\documentclass[]{article}
\usepackage{url} 
\usepackage{hyperref}
\hypersetup{colorlinks, linkcolor=blue, citecolor=blue, urlcolor=blue}

\usepackage{lineno}
\usepackage{soul}
\usepackage{tikz}
\usetikzlibrary{shapes, arrows}
\usetikzlibrary{cd}
\usetikzlibrary{trees,arrows.meta,shapes.geometric,decorations.pathreplacing,calc}
\usetikzlibrary{positioning, arrows.meta}
\usepackage[ruled,vlined]{algorithm2e} 

\usepackage[dvipsnames]{xcolor}

\tikzstyle{terminator} = [rectangle, draw, text centered, rounded corners, minimum height=2em]
\tikzstyle{process} = [rectangle, draw, text centered, minimum height=2em]
\tikzstyle{decision} = [diamond, draw, text centered, minimum height=2em]
\tikzstyle{data}=[trapezium, draw, text centered, trapezium left angle=60, trapezium right angle=120, minimum height=2em]
\tikzstyle{connector} = [draw, -latex']
\tikzset{curve/.style={settings={#1},to path={(\tikztostart)
    .. controls ($(\tikztostart)!\pv{pos}!(\tikztotarget)!\pv{height}!270:(\tikztotarget)$)
    and ($(\tikztostart)!1-\pv{pos}!(\tikztotarget)!\pv{height}!270:(\tikztotarget)$)
    .. (\tikztotarget)\tikztonodes}},
    settings/.code={\tikzset{quiver/.cd,#1}
        \def\pv##1{\pgfkeysvalueof{/tikz/quiver/##1}}},
    quiver/.cd,pos/.initial=0.35,height/.initial=0}
    
\newcommand{\ra}[1]{\renewcommand{\arraystretch}{#1}}
\usepackage{amsmath,amssymb,amsfonts}%
\usepackage{amsthm}%
\usepackage{float}
\usepackage{soul}
\usepackage{subcaption}
\usepackage{authblk}
\usepackage{apacite}

\definecolor{pygreen}{RGB}{1,128,0}    
\definecolor{pypurple}{RGB}{128,0,128} 
\definecolor{pyorange}{RGB}{250,164,4}  
\definecolor{pyblue}{RGB}{0,1,241}    

\usepackage{geometry}[margin = 2.54cm]

\begin{document}

\title{Learning vertical coordinates via automatic differentiation of a dynamical core}

\author[1]{Tim Whittaker\thanks{Corresponding author: whittaker.tim@courrier.uqam.ca}}
\author[2]{Seth Taylor}
\author[3]{Elsa Cardoso-Bihlo}
\author[1]{Alejandro Di Luca}
\author[3]{Alex Bihlo}

\affil[1]{Centre pour l'étude et la simulation du climat à l'échelle régionale (ESCER), D\'epartement des Sciences de la Terre et de l’Atmosph\`ere, Universit\'e du Qu\'ebec \`a Montr\'eal, Montr\'eal,  Qu\'ebec, Canada}
\affil[2]{PIMS, Department of Computer Science, University of Saskatchewan, Saskatoon, Canada}
\affil[3]{Department of Mathematics and Statistics, Memorial University of Newfoundland, St. John's, Canada}

\maketitle

\begin{abstract}
Terrain-following coordinates in atmospheric models often imprint their grid structure onto the solution, particularly over steep topography, where distorted coordinate layers can generate spurious horizontal and vertical motion. Standard formulations, such as hybrid or SLEVE coordinates, mitigate these errors by using analytic decay functions controlled by heuristic scale parameters that are typically tuned by hand and fixed a priori. In this work, we propose a framework to define a parametric vertical coordinate system as a learnable component within a differentiable dynamical core. We develop an end-to-end differentiable numerical solver for the two-dimensional non-hydrostatic Euler equations on an Arakawa C-grid, and introduce a NEUral Vertical Enhancement (NEUVE) terrain-following coordinate based on an integral transformed neural network that guarantees monotonicity. A key feature of our approach is the use of automatic differentiation to compute exact geometric metric terms, thereby eliminating truncation errors associated with finite-difference coordinate derivatives. By coupling simulation errors through the time integration to the parameterization, our formulation finds a grid structure optimized for both the underlying physics and numerics. Using several standard tests, we demonstrate that these learned coordinates reduce the mean squared error by a factor of 1.4 to 2 in non-linear statistical benchmarks, and eliminate spurious vertical velocity striations over steep topography.
\end{abstract}


\newpage
\section{Introduction}
\label{intro}

High-resolution numerical weather prediction and climate models are essential tools for capturing the non-linear, multiscale interactions between atmospheric dynamics and complex topography \cite{Bauer2015,roberts_high-resolution_2025,haarsma_high_2016}. As the modeling community progresses toward convection-permitting resolutions ($<4$ km), the fidelity of the lower boundary condition becomes increasingly critical \cite{Bauer2015, Wedi2025}. At these scales, models must explicitly resolve steep orographic features, since the interaction of the orography with the lower atmosphere generates a vertical flow that requires solving a system of non-hydrostatic equations. This change in dynamics leads to the generation of complex pressure perturbations, vertical momentum flux, and buoyancy-driven instability near the surface. These features, when resolved,
mechanically force gravity waves, drive orographic precipitation, and govern local circulation patterns \cite{Kirshbaum2018, Rotach2014}. To accommodate irregular lower boundaries, the vast majority of atmospheric models employ terrain-following coordinates. \par

The classical sigma ($\sigma$) coordinate system \cite{ACOORDINATESYSTEMHAVINGSOMESPECIALADVANTAGESFORNUMERICALFORECASTING} maps the physical domain to a rectangular computational grid, simplifying the application of boundary conditions. This transformation introduces a well-documented source of discretization error: the calculation of the horizontal pressure-gradient force requires the cancellation of two large terms, and over steep slopes, truncation errors in this balance can produce spurious accelerations comparable in magnitude to the physical flow \cite{Mesinger01021982, ANewTerrainFollowingVerticalCoordinateFormulationforAtmosphericPredictionModels}. As model resolution increases, resolved topographic slopes become steeper, amplifying these errors and threatening the stability of the model. \par

Standard mitigation strategies rely on hybrid vertical coordinates \cite{Simmons1981} that transition from terrain-following layers near the surface to quasi-horizontal surfaces in the free atmosphere. In practice, ensuring stability over steep slopes often necessitates the pre-filtering of the underlying topography \cite{https://doi.org/10.1029/2019MS001781}, a compromise that reduces numerical noise but indiscriminately removes resolved-scale orographic forcing \cite{Elvidge2019}. Advanced formulations, such as the SLEVE coordinate \cite{ANewTerrainFollowingVerticalCoordinateFormulationforAtmosphericPredictionModels} or smoothed-level coordinates \cite{ATerrainFollowingCoordinatewithSmoothedCoordinateSurfaces}, achieve this mitigation by applying analytic decay functions to attenuate small-scale topographic signatures with height. Consequently, the majority of solvers continue to rely on analytic coordinate formulations with heuristic scale-height parameters fixed a priori~\cite{beck2020evaluation}. 

The emerging paradigm of differentiable programming offers a pathway to overcome these rigid architectural choices \cite{sapienza2024differentiableprogrammingdifferentialequations}. By constructing ``end-to-end'' differentiable dynamical cores, we can utilize automatic differentiation (AD) to optimize numerical components that were previously hand-tuned \cite{gmd-16-3123-2023}. Recent work has demonstrated the efficacy of this approach for objective calibration and hybrid physics-ML parameterizations \cite{Kochkov2024, gmd-18-3017-2025}. In this work, we extend this methodology to account for the discretization itself and propose a terrain-following coordinate defined through optimization of a differentiable dynamical core. \par

The field of numerical weather prediction is witnessing a paradigm shift driven by deep learning~\cite{BAUER2024100002}. Purely data-driven ML weather forecasting models have recently demonstrated predictive skill competitive with state-of-the-art operational systems, offering orders-of-magnitude acceleration in computational throughput \cite{Bi2023, Lam2023}. Although these surrogate models excel at short-term prediction within observed climate distributions \cite{https://doi.org/10.1029/2023MS004019} they act as statistical emulators rather than physical solvers~\cite{lai2025machine}. As a result, they do not strictly enforce conservation laws (such as mass, momentum, energy, or potential vorticity), leading to physical inconsistencies, spectral blurring, and the loss of fine-scale detail over long integration times. Recent studies indicate that these limitations can affect the representation of extreme weather events \cite{Pasche2025, Zhang2025} and reduce reliability in out-of-distribution climate regimes \cite{Sun2025}. Furthermore, the reliability of pure ML models in the high-resolution convection-permitting regimes targeted by regional climate models (RCMs) remains an open question. While recent work has attempted to bridge this gap \cite{pathak2024kilometerscaleconvectionallowingmodel}, the majority of approaches rely on static downscaling or super-resolution rather than temporal forecasting \cite{Harder2022, Mardani2025}. These limitations create a compelling motivation for hybrid frameworks such as NeuralGCM \cite{Kochkov2024}, which fuse the physical modeling of traditional dynamical cores with the optimization capabilities of machine learning (ML). Our work also adopts this hybrid philosophy: rather than replacing the governing equations with a black box, we utilize differentiable programming to optimize the numerical discretization itself.

We present a differentiable non-hydrostatic dynamical core with a differentiable parametric vertical coordinate transformation. By propagating gradients through the solver, the coordinate system can be optimized to minimize simulation error, effectively discovering a hybrid grid structure for the underlying physics. To validate this approach, we adopt the benchmark suite inspired by \citeA{GIRALDO20083849}. 

The work is organized as follows: In Section~\ref{Sec:Methods}, we detail the governing non-hydrostatic Euler equations, the analytic and neural vertical coordinate formulations, and the gradient-based optimization framework. Section~\ref{Sec:Results} presents the comparative results across the hierarchy of test cases, ranging from linear advection to the non-linear density current, and analyzes the geometric structure of the learned grid. In Section~\ref{Sec:Discussion}, we discuss the generalization capabilities of the method, the implications for operational modeling, and the advantages of exact metric computation via automatic differentiation. Finally, Section~\ref{Sec:Conclusions} summarizes our findings and outlines future research directions.

\section{Methods}
\label{Sec:Methods}

In this section, we define the governing physical equations, the discrete grid formulation, the neural network parameterization of the vertical coordinate, and the coordinate optimization routine.

\subsection{Governing equations}

To simulate atmospheric dynamics, we solve the two-dimensional compressible Euler equations in a vertical slice $\mathbf{x}=(x,z)$ and  adopt the non-conservative formulation analysed by \citeA{GIRALDO20083849} (Equation Set 1). This utilizes the wind field $\mathbf{u}=(u,w)^{\rm T}$, the Exner pressure function $\pi=(p/p_0)^{R/c_p}$ and potential temperature $\theta$ as prognostic variables with governing dynamics 
\begin{align}\label{eq:EulerEquations}
\begin{split}
&\frac{\partial \mathbf{u}}{\partial t} + \mathbf{u} \cdot \nabla \mathbf{u} + c_p \theta \nabla \pi = -g \mathbf{k} + \nu\nabla^2\mathbf{u}, \\
&\frac{\partial \pi}{\partial t} + \mathbf{u} \cdot \nabla \pi + \frac{R}{c_v} \pi \nabla \cdot \mathbf{u} = 0, \\
&\frac{\partial \theta}{\partial t} + \mathbf{u} \cdot \nabla \theta = 0,
\end{split}
\end{align}
where $\nabla$ is the gradient in $(x,z)$, $g$ is gravitational acceleration, $\mathbf{k}$ is the vertical unit vector, and $c_p, c_v, R$ are the standard thermodynamic constants for dry air. We note that for the standard Euler equations, the diffusion coefficient $\nu$ is zero, but in some of the test cases by~\citeA{GIRALDO20083849} physical diffusion is explicitly added, thus effectively solving the Navier-Stokes equations. The momentum equation couples the thermodynamic state to the grid geometry through the pressure gradient term $c_p \theta \nabla \pi$.

\subsection{Vertical coordinate formulations}\label{subsec:VerticalCoodinates}

Since meteorological models demand realistic bottom topography, using the standard Cartesian vertical coordinate $z$ is problematic because its surfaces intersect the terrain, necessitating grid point masking and introducing challenging boundary conditions. Instead, one traditionally employs terrain-following coordinate systems that map the physical height $z$ to a computational vertical coordinate $\zeta$. The bottom topography then naturally becomes the lower boundary condition of the new vertical coordinate, and the transformed coordinate is defined by
\begin{equation}
\label{eq:transformation}
z(x, \zeta) = \zeta + h(x) \cdot B(\zeta),
\end{equation}
where $B(\zeta)$ is a monotonically decreasing vertical decay function satisfying $B(0)=1$ (terrain-following lower boundary) and $B(H)=0$ (flat top of atmosphere boundary). The definition of $B(\zeta)$ determines how rapidly the influence of topography attenuates with height, which directly impacts the magnitude of coordinate-induced errors. The Euler equations~\eqref{eq:EulerEquations} are then transformed into the new coordinate system $(x,\zeta)$, which introduces extra metric terms described in Appendix~\ref{app:derivation}.

\begin{figure}[h!]
    \centering
    \includegraphics[width=\linewidth]{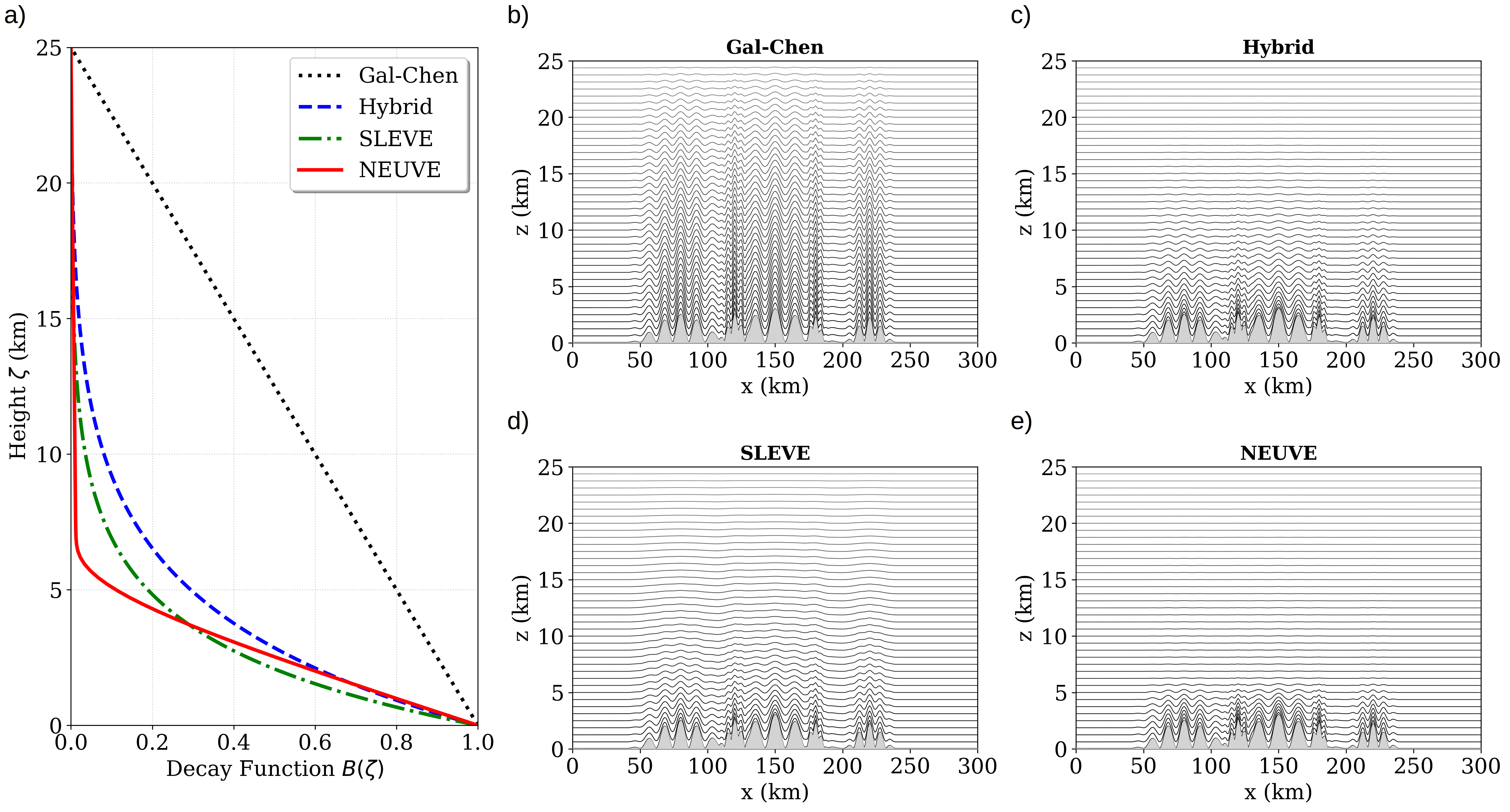}
    \caption{(a) Vertical decay functions $B(\zeta)$ for Gal-Chen and Somerville (dotted), Hybrid (dashed), SLEVE (dash-dot), and the learned NEUVE (solid red). (b-e) Visual comparison of the resulting computational grids over a sample mountain profile for (b) Gal-Chen and Somerville, (c) Hybrid, (d) SLEVE, and (e) NEUVE coordinates.}
    \label{fig:topo_comp}
\end{figure}

In this work, we consider four formulations of the decay function $B(\zeta)$ which are depicted in Figure \ref{fig:topo_comp}. The analytic definition for each transformation is given by
\begin{subequations}
\begin{align}
\text{Gal-Chen and Somerville}:&\quad  B(\zeta) = 1- \zeta/H \,, \label{coord:gal_chen}
\\
\text{Hybrid}:& \quad B(\zeta) = (1 - \zeta/H) \exp(-\zeta / s)\,, \label{coord:hybrid}
\\
\text{SLEVE}:& \quad B(\zeta) = \sum_{i = 1}^2h_i \sinh[(H/s_i) - (\zeta/s_i)] / \sinh[H/s_i]\,, \label{coord:Sleve}
\\
\text{NEUVE}: & \quad B(\zeta) = \frac{1}{N(\phi)}\int_0^{\zeta/H} \rho_{\phi}(s) \mathrm{d}s\,. \label{coord:Neuve}
\end{align}
\end{subequations}
The Gal-Chen and Somerville coordinate \eqref{coord:gal_chen} was introduced by \citeA{gal1975use} and contains no tunable parameters. The Hybrid coordinate \eqref{coord:hybrid} was proposed by \citeA{Simmons1981} and employs an exponential decay controlled by a scale height parameter $s$. The smooth level vertical (SLEVE) coordinate \eqref{coord:Sleve} decomposes the topography into large-scale ($s_1$) and small-scale ($s_2$) components following the work by \citeA{ANewTerrainFollowingVerticalCoordinateFormulationforAtmosphericPredictionModels}, applying scale-dependent decay functions to rapidly smooth small-scale features that generate the largest numerical errors.\par
Our proposed\textit{NEUVE (NEUral Vertical Enhancement)} coordinate utilizes a decay function which is parameterized by a neural network where $\phi$ are the trainable weights and biases. We utilize the composite function 
\begin{equation}
    \rho_{\phi}(\eta) = \log(1 + \exp(f_{\phi}(\eta))) + \epsilon,
\end{equation}
which is the Softplus activation function with a small constant shift to ensure strict positivity. In our numerical results we used $\epsilon=0.05$. The decay profile $B(\zeta)$ is then recovered via numerical integration to enforce strict monotonicity, preventing coordinate surfaces from intersecting. The coordinate \eqref{coord:Neuve} satisfies $B(0)=1$, $B(H)=0$, and $\partial B / \partial \zeta < 0$ for all $\zeta$ by construction ensuring a valid terrain-following coordinate system regardless of the network weights. The function $f_{\boldsymbol{\phi}}$ is taken to be a Multi-Layer Perceptron (MLP) with hyperbolic tangent activations. Using a hyperparameter sensitivity analysis, detailed in Appendix \ref{APP:Hyper}, we found an optimal configuration with 3 dense layers of width 64. The other training parameters are also specified in Appendix \ref{APP:Hyper}.\par

\subsection{Numerical discretization}

The governing equations are discretized in $(x,\zeta)$ coordinates on an Arakawa C-grid \cite{Arakawa1977}, where scalar variables ($\pi, \theta$) are defined at cell centers, while velocity components are staggered at cell faces. Spatial derivatives are approximated using second-order centered finite differences and the time-stepping is performed with the third order Runge-Kutta Strong-Stability-Preserving scheme. The solver uses a hyperdiffusion in the prognostic momentum equations for stability of the explicit time-stepping. Our implementation is written in \texttt{JAX}, which is a high-performance Python library for ML and scientific computing that enables automatic differentiation of native Python functions~\cite{jax2018github}. Discretizing the Euler equations in a generalized vertical coordinate involves metrics arising from the Jacobian of the transformation which are commonly approximated using finite differences. This can introduce geometric truncation errors and is prone to generating spurious numerical results, cf.~\citeA{klemp2003numerical}. We instead base our differentiable numerical solver on the exact derivatives of the coordinate transformation functions $z(\xi, \zeta)$ which can be computed by hand or via automatic differentiation. This ensures that the geometric terms are accurate to machine precision, isolating the discretization error in the coordinate optimization routines from the error induced by the transformation.  We provide further technical details on the numerical discretization of system~\eqref{eq:EulerEquations} and a validation of our numerical schemes in Appendix \ref{app:numerics} while specific experimental parameters are tabulated in Appendix \ref{APP:Exp_params}.\par

\subsection{Coordinate optimization routines}
\label{subsec:optimization}

We define the vertical coordinate transformation as the solution to a constrained optimization problem. The objective functional is constructed to minimize the numerical discretization error while the constraint is impose to ensure numerical stability. We use a soft-constrained regularization and perform the optimization using a \textit{solver-in-the-loop} training procedure~\cite{um2020solver}: at each iteration we generate a coordinate mapping, run the forward model, and compute the objective which respect to the coordinate transformation parameters through the solver steps. Denoting $\phi$ as the parameters of the coordinate transformation and $\psi_t(\cdot,\phi): \mathbb{R}^N \to \mathbb{R}^N$ to be the forward model computed as the composition of time steps
\begin{equation}
\psi_t(\mathbf{x}_i,\phi) = \psi_{\Delta t}(\cdot,\phi)\circ \psi_{\Delta t}(\cdot,\phi)\circ  \cdots \circ \psi_{\Delta t}(\mathbf{x}_i,\phi)
\end{equation}
with $t = N_t\Delta t$, the objective functional can be written generically in the form
\begin{equation}\label{eq:LossFunction}
    \mathcal{J}(\phi) = \mathcal{L}_{\rm physics}(\psi_t(\phi)) + \lambda \mathcal{L}_{\rm reg}(\phi).
\end{equation}
The regularization should be sufficient to maintain stability of the solver when applied across the dynamical scenarios of interest, whereas the definition of the physics objective can target a particular dynamical scenario or topography. Part of the benefit of the coordinate systems Eqn. \eqref{coord:gal_chen}, Eqn. \eqref{coord:Sleve}, Eqn. \eqref{coord:hybrid} is that they can be applied across different dynamical regimes, solvers, and bottom topographies. For this perspective, we define a physics loss in terms of an expectation proxy error over a distribution of topographies. By randomizing the domains, we expose the solver to a diverse set of possible geometric configurations during training leading to more robust learned decay profiles. In order to expose the numerical truncation error as much as possible, we consider training on a reference solution defined by the transport of a scalar density $\mu$ in a prescribed velocity field:
\begin{equation}
\label{eq:advection}
\frac{\partial \mu}{\partial t} + \nabla \cdot (\mu \mathbf{u}) = 0.
\end{equation}
The physics loss is defined as the Root Mean Squared Error (RMSE) of the prognostic variable (scalar quantity $\mu$ for advection tests) against the analytic reference solution $\psi_{ref}$:
\begin{equation}
    \mathcal{L}_{\rm physics} = \left(\frac{1}{N} \sum_{i=1}^{N} || \psi_{\rm model}(\mathbf{x}_i, T; \phi) - \psi_{\rm ref}(\mathbf{x}_i, T) ||^2\right)^{1/2}.
\end{equation}
This procedure is summarized in Algorithm~\ref{alg:training_loop}.

\begin{algorithm}[H]
\caption{Coordinate optimization routine}
\label{alg:training_loop}
\DontPrintSemicolon
\KwIn{Training distribution $\mathcal{M}(\mathcal{S})$, horizon $T$, timestep $\Delta t$, initial parameter $\phi^{(0)}$.}
\KwOut{Updated parameters $\phi^*$.}

\For{$k = 0,1,2,\dots$}{
  \tcp{Topography sampling}
    Sample $\boldsymbol{\tau} \sim \mathcal{M}(\mathcal{S})$ and synthesize bottom topography $h(x;\boldsymbol{\tau})$.\\
    \vspace{0.3cm}
    
  \tcp{Coordinate generation}
    Compute vertical decay profile $B(\zeta;\phi^{(k)})$ and construct $z(x,\zeta;\phi^{(k)}) \gets (h,B)$ using \eqref{eq:transformation}.\\
    Precompute metric terms from $z_{\xi}$ and $z_{\eta}$ \\
    \vspace{0.3cm}

  \tcp{Forward model solve}
  Run the numerical solver in the transformed domain up to time $T$ to obtain $\psi_T(\phi^{(k)})$ .\\
    \vspace{0.3cm}

  \tcp{Loss evaluation and parameter update}
  Compute gradients $g^{(k)} \gets \nabla_{\phi}(\mathcal{L}_{\rm physics}(\psi_T(\phi^{(k)})) + \lambda_{\rm reg}\,\mathcal{L}_{\rm reg}(\phi^{(k)}))$.
  \\
  Update $\phi^{(k+1)} \gets (\phi^{(k)}, g^{(k)})$ using Adam.\\[2pt]
}
\Return{$\phi^* = \phi^{(k)}$ upon convergence.}
\end{algorithm}

\subsection{Implementation details}

The proposed framework presents several algorithmic choices. Below, we detail the specific selections made for this study, their rationale, and the alternatives available within the methodology.\\

\noindent
\emph{Proxy training task:} We chose to optimize the coordinate transformation on the linear advection Eqn.~\eqref{eq:advection}. We selected this task because the availability of an analytical solution provides a precise training target $\psi_{\rm ref}$ for the loss function, as well as having been used as the benchmark for defining the SLEVE coordinate~\cite{ANewTerrainFollowingVerticalCoordinateFormulationforAtmosphericPredictionModels}. An alternative approach, which would not require a reference solution, would be to define a loss function based directly on theoretical discretization error estimates, such as those derived by \citeA{ANewTerrainFollowingVerticalCoordinateFormulationforAtmosphericPredictionModels}.\\

\noindent
\emph{Domain randomization:} We utilized a domain randomization strategy where the topography is generated at each training distribution by sampling from a distribution $\mathcal{M}(\mathcal{S})$. The topographies $h(x;\mathbf{\tau})$ are defined parametrically as a large-scale bell-shaped mountain modulated by small-scale oscillations:
\begin{equation}
    \label{eq:topo}
    h(x;\mathbf{\tau}) = \begin{cases}
    h_0 \cos^2\left(\frac{\pi x}{2a_c}\right) \cos^2\left(\frac{\pi x}{\lambda}\right) & \text{for } |x| \le a_c \\
    0 & \text{otherwise}.
    \end{cases}
\end{equation}
with height $h_c$, half-width $a_c$, and small-scale wavelength $\lambda_c$. The height and center of each mountain were sampled from global uniform distributions, with $h_c \in [500, 3000]$~m and $x_c \in [50, 250]$~km. The shape parameters (half-width $a_c$ and wavelength $\lambda_c$) were sampled conditionally from three equally probable regimes ($p=1/3$):
\begin{itemize}
    \item \textit{Smooth}: $a_c \in [40, 80]$~km, $\lambda_c \in [12, 25]$~km;
    \item \textit{Standard}: $a_c \in [15, 40]$~km, $\lambda_c \in [8, 12]$~km;
    \item \textit{Jagged}: $a_c \in [8, 20]$~km, $\lambda_c \in [5, 9]$~km.
\end{itemize}
This strategy exposes the network to high-frequency, jagged terrain features during optimization, preventing overfitting to smooth, idealized geometries. However, this randomization is a design choice; the framework can equally be applied to optimize a coordinate system for a single, fixed topography, or extended to randomize the initial flow conditions (e.g., varying the shear layer height) to prevent overfitting to a specific wind profile.\\

\noindent
\emph{Coordinate parameterization:} We parameterized the vertical decay function $B(\zeta)$ using NEUVE to allow for the discovery of non-analytic vertical structures. Alternatively, the differentiable framework can be applied to standard analytic formulations. In Appendix \ref{app:param_opt}, we demonstrate how the same optimization loop can be used to tune scalar parameters (e.g., scale heights $s_1, s_2, s$) of the SLEVE or Hybrid coordinate. To evaluate the integral in Eqn.~\eqref{coord:Neuve} efficiently during training, we approximate the continuous integral using a cumulative Riemann sum on a fixed grid. We utilize 100 equidistant intervals spanning the normalized vertical domain $[0, 1]$.\\

\noindent
\emph{Stability regularization:} To strictly enforce grid monotonicity during the early phases of optimization, a ReLU-based penalty is applied to the coordinate Jacobian $J = \partial z / \partial \zeta$. We selected a regularization weight of $\lambda_{\rm reg} = 10^{-7}$ to penalize invalid grid folding. Furthermore, a conditional crash recovery mechanism bypasses the gradient update if the simulation produces non-finite values or if the Jacobian falls below a critical threshold ($J_{\rm min} < 10^{-5}$). To actively discourage grid tangling during the early phases of training, we apply a regularization penalty to the coordinate Jacobian $J = \partial z / \partial \zeta$:
\begin{equation}
    \mathcal{L}_{\rm reg} = \sum \text{ReLU}(-J),
\end{equation}
with weighting $\lambda_{\rm reg} = 10^{-7}$, where $\mathrm{ReLU}=\max(0,x)$ is the rectified linear unit. Finally, to prevent invalid gradients from corrupting the network weights during divergent simulations, we implement a conditional crash recovery mechanism. If the minimum Jacobian falls below a critical threshold ($J_{\rm min} < 10^{-5}$) or if the simulation produces NaNs, the physics loss is bypassed and replaced with a fixed high penalty ($\mathcal{L} = 1.0$). \\

\noindent
\emph{Loss minimization:} We use the standard Adam optimizer~\cite{kingma2014adam} to minimize the loss function~\eqref{eq:LossFunction}. The learning rate $\eta$ is set to $\eta=10^{-3}$ and loss gradients are clipped to ensure the $L_2$-norm of gradients cannot exceed 1. 

\newpage
\section{Results}
\label{Sec:Results}

The coordinate optimization framework is validated using a hierarchy of test cases adapted from standard non-hydrostatic benchmarks proposed by \citeA{GIRALDO20083849}, which range from idealized linear advection to highly non-linear density currents. The parameters used for each simulation are recorded in Table~\ref{sim_details}. 

\subsection{Case 1: Advection of a passive scalar}

We begin with the idealized two-dimensional advection test defined by \citeA{ANewTerrainFollowingVerticalCoordinateFormulationforAtmosphericPredictionModels}. The velocity field of this test case is chosen as $\mathbf{u}=(u(z), 0)$. The horizontal wind profile $u(z)$ is specified to be zero near the surface (a stagnant layer) and transitions to a constant background flow aloft through a shear layer:
\begin{equation}
    \label{eq:shar_profile}
    u(z) = \begin{cases}
        0 & \text{for } z \le z_1 \\
        u_0 \sin^2\left(\frac{\pi}{2} \frac{z - z_1}{z_2 - z_1}\right) & \text{for } z_1 < z < z_2 \\
        u_0 & \text{for } z \ge z_2,
    \end{cases}
\end{equation}
where $u_0 = 10$ ms$^{-1}$ is the maximum velocity, $z_1 = 4$ km is the height of the stagnant layer, and $z_2 = 5$ km is the base of the constant flow layer.

A cosine-shaped scalar anomaly is initialized upstream at $(x_c, z_c) = (-50 \text{ km}, 9 \text{ km})$ and advected over the mountain. In a perfect discretization, the shape of the anomaly is preserved; however, coordinate distortions typically introduce dispersive errors. This case serves as the initial training objective for the neural network parameterization.

\begin{figure}[h!]
    \centering
    \includegraphics[width=0.75\linewidth]{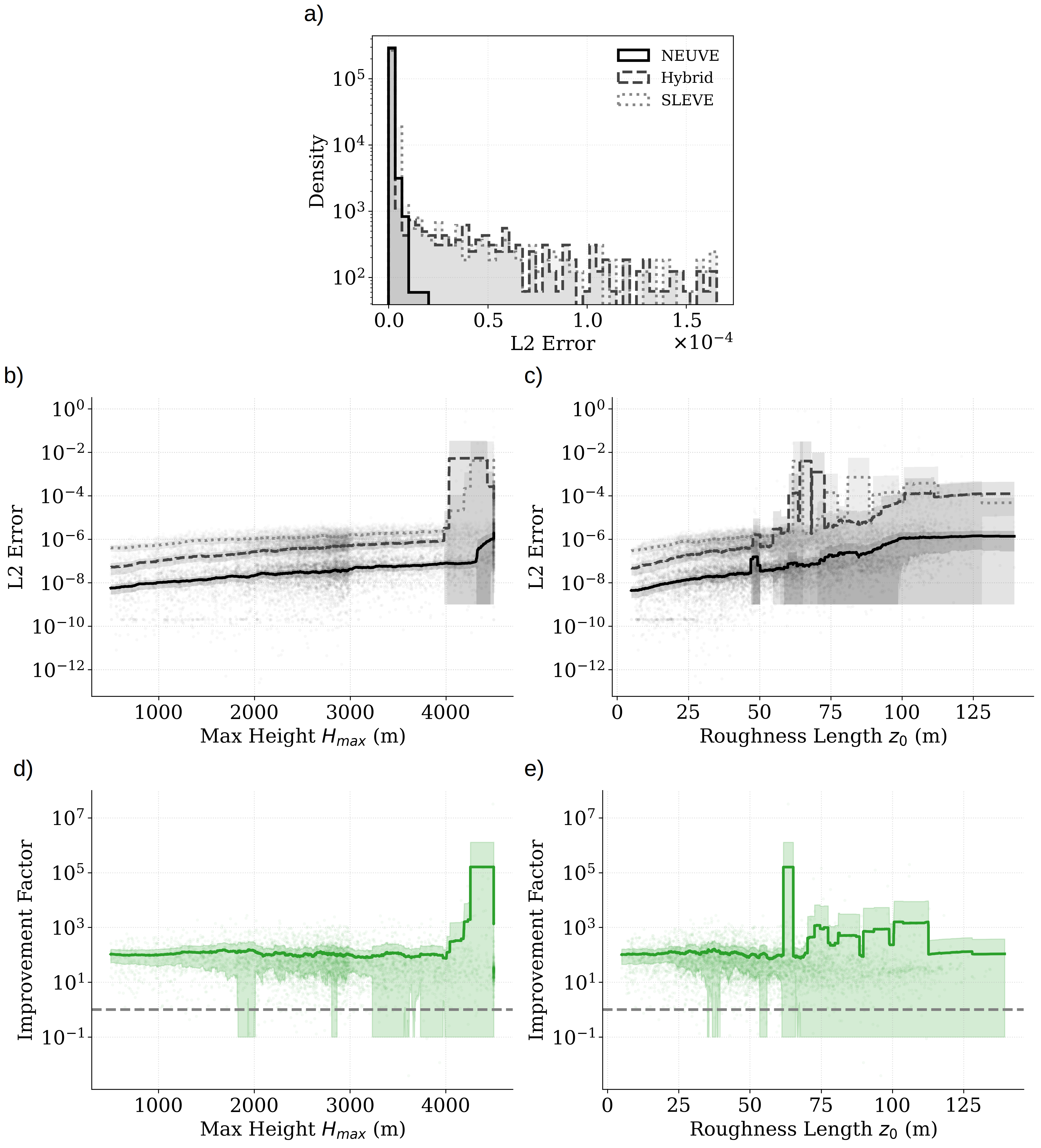}
    \caption{Statistical evaluation of coordinate performance on the passive scalar advection test case. (a) Probability density of $L_2$ errors across the stochastic topography test ensemble for NEUVE (solid), Hybrid (dashed), and SLEVE (dotted) coordinates. (b) Sensitivity of $L_2$ error to maximum mountain height ($H_{max}$). (c) Sensitivity of $L_2$ error to topographic roughness length ($z_0$). Shaded regions in (b) and (c) indicate the error variance across the batch. (d--e) The improvement factor, defined as the ratio of the SLEVE $L_2$ error to NEUVE $L_2$ error, plotted against (d) $H_{max}$ and (e) $z_0$. The classic Gal-Chen and Somerville coordinate is excluded from these comparisons as its errors are orders of magnitude larger, which would distort the graphical scale.}
    \label{fig:advect_stats}
\end{figure}

The neural network coordinate parameterization NEUVE was trained exclusively on the linear advection Eqn. ~\eqref{eq:advection} using the stochastic topography generation protocol described in Section~\ref{subsec:optimization}. Figure~\ref{fig:advect_stats} summarizes the statistical performance of the NEUVE coordinate against the standard Hybrid, and SLEVE formulations across an ensemble of unseen test topographies. The Gal-Chen and Somerville coordinate was not included here since it causes much larger errors than the other schemes, which would distort the graphical scale.

\begin{figure}[h!]
    \centering
    \includegraphics[width=0.75\linewidth]{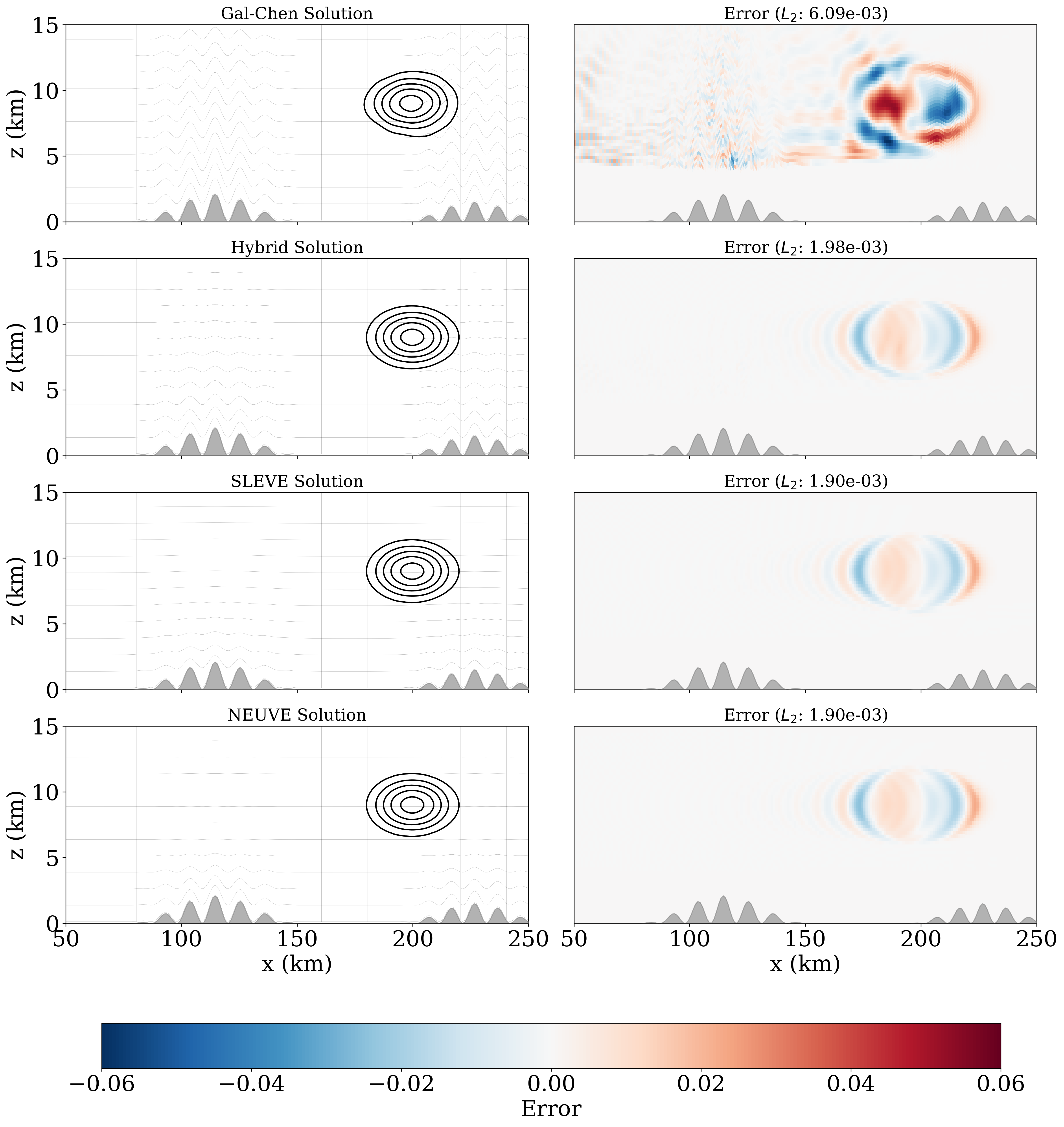}
    \caption{Visual comparison of advection accuracy across coordinate formulations for a randomized topography sample within the extra advection test case. The left column displays the final state ($t=T$) of the passive scalar contour after advecting over a rough surface profile, while the right column shows the corresponding error field defined as the difference between the model and the exact solution. The rows, from top to bottom, correspond to the Gal-Chen and Somerville, Hybrid, SLEVE, and NEUVE coordinate solutions.}
    \label{fig:advection_test}
\end{figure}

The distribution of $L_2$-errors depicted in Fig.~\ref{fig:advect_stats}a shows the NEUVE coordinate significantly shifts the error density toward zero, eliminating the heavy tail of high-error outliers observed in the Hybrid and SLEVE formulations. To isolate the regimes where the NEUVE coordinate offers distinct advantages, we analyze the error scaling with respect to topographic characteristics. While all coordinate systems exhibit increased error with maximum mountain height ($H_{max}$), the NEUVE coordinate maintains the lowest error magnitude throughout the tested range, see Fig.~\ref{fig:advect_stats}b.

The performance divergence is also present with respect to topographic roughness ($z_0$), defined here as the standard deviation of the terrain height normalized by a constant (Fig.~\ref{fig:advect_stats}c). As the roughness increases—representing high-frequency, jagged terrain features that induce rapid grid curvature—the errors for all the schemes increased though the NEUVE coordinate have an overall lower error. 

This advantage is quantified by the improvement factor, defined as the ratio of the $L_2$-error of the SLEVE coordinate to that of the NEUVE coordinate. While the improvement remains consistent across varying mountain heights (Fig.~\ref{fig:advect_stats}d), it exhibits a distinct regime shift with respect to roughness (Fig.~\ref{fig:advect_stats}e). As $z_0$ increases beyond 75 m, the improvement factor drops but stabilize. Figure~\ref{fig:advection_test} shows specific examples of the solutions and errors for a given topography. The distortion of the solution from the vertical coordinate is the most noticeable for the Gal-Chen and Somerville coordinate while the other coordinates show similar error.

\subsection{What coordinates have we learned?}

To understand the geometric strategy driving the performance gains observed in Case 1, we examine the vertical decay functions $B(\zeta)$ derived from the optimization process. Figure \ref{fig:topo_comp}a compares the learned NEUVE profile against the standard analytic formulations.

The neural network converges to a grid structure that differs from the heuristic exponential or scale-dependent decays of the Hybrid and SLEVE coordinates. While the analytic formulations distribute the topographic influence gradually throughout the column, the learned profile identifies a distinct two-regime strategy. It maintains a strict terrain-following structure near the surface but transitions sharply to a flat grid ($B(\zeta) \approx 0$) at an altitude of approximately $z=5$ km.

This transition altitude coincides exactly with the base of the constant flow layer ($z_2 = 5$ km) defined in the advection wind profile given through Eqn.~\eqref{eq:shar_profile}. This implies that the optimization process implicitly identified the shear layer ($z_1 < z < z_2$) as the region for discretization error. By driving the coordinate transformation to zero at $z_2$, the network minimizes the geometric truncation errors associated with calculating horizontal fluxes across distorted coordinate surfaces within the constant flow layer.

The resulting spatial grids are visualized in Figures \ref{fig:topo_comp}b--e. The Gal-Chen and Somerville coordinate (Fig.~\ref{fig:topo_comp}b) transmits topographic undulations to the model top, creating skewed cells throughout the domain. The Hybrid and SLEVE grids (Fig.~\ref{fig:topo_comp}c, d) attenuate these undulations but retain non-zero curvature at mid-levels. In contrast, the NEUVE coordinate (Fig.~\ref{fig:topo_comp}e) creates a perfectly Cartesian mesh above 5 km. This suggests that for this specific flow configuration, the solver minimizes error not by maximizing smoothness alone, but by maximizing orthogonality in regions of strong transport. However, this sharp adaptation to the specific vertical level $z_2$ indicates that the network has optimized the grid geometry specifically for the vertical structure of the training flow.

\subsection{Case 2: Rising bubble}

To evaluate the coordinate system in a regime dominated by strong vertical motion, we employ a rising thermal bubble test adapted from the standard benchmark of \citeA{GIRALDO20083849}. 

The background atmosphere is initialized in hydrostatic balance with a constant potential temperature $\theta_0 = 300$ K (neutral stratification). A warm perturbation is introduced in the center of the domain at $(x_c, z_c) = (500, 350)$ m:
\begin{equation}
    \Delta \theta(x,z) = 
    \begin{cases} 
      \Delta \theta_c \cos^2 \left( \frac{\pi r}{2R} \right) & \text{for } r \le R \\
      0 & \text{otherwise},
    \end{cases}
\end{equation}
where $r = \sqrt{(x-x_c)^2 + (z-z_c)^2}$ is the radial distance from the perturbation center. We utilize a perturbation amplitude of $\Delta \theta_c = 0.5$ K and a radius $R = 250$ m. The topography follows Eqn.~\eqref{eq:topo}. To generate the test ensemble, we sample the topographic parameters $h_c$, $a_c$, and $\lambda_c$ from random distributions. However, because many such random configurations create steep gradients that cause numerical instabilities in the standard Hybrid coordinate, we employ a rejection sampling strategy. We filter the ensemble to include only those topographic profiles where all coordinate systems successfully complete the simulation without instability.

By initializing the thermal perturbation sufficiently high above the topography (ensuring the perturbation radius $R$ does not overlap with the maximum topographic height $h_{\max}$), the physical solution remains largely independent of the terrain. While the lower boundary induces minor far-field flow adjustments to the bubble's return circulation, these effects are negligible compared to the numerical imprint of the grid structure. Consequently, any significant divergence from the flat-terrain benchmark can be effectively isolated as a coordinate-induced discretization error.

\begin{figure}[h!]
    \centering
    \includegraphics[width=0.75\linewidth]{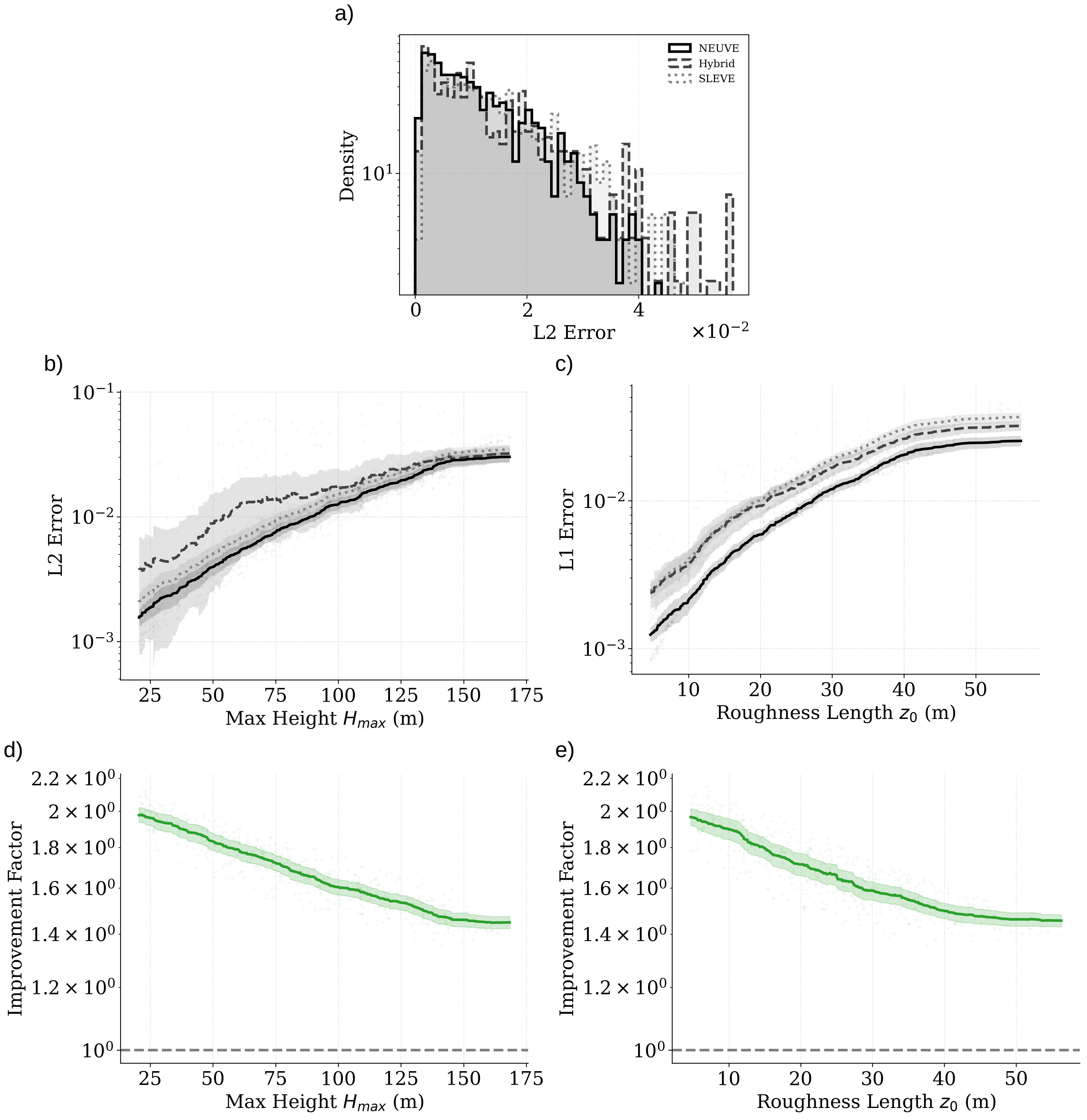}
    \caption{Statistical evaluation of coordinate performance on the rising thermal bubble test case. (a) Probability density of $L_2$-errors across the stochastic topography test ensemble ($N=500$) for NEUVE (solid), Hybrid (dashed), and SLEVE (dotted) coordinates. (b) Sensitivity of $L_2$-error to maximum mountain height ($H_{max}$). (c) Sensitivity of $L_2$-error to topographic roughness length ($z_0$). The improvement factor, defined as the ratio of the SLEVE $L_2$-error to NEUVE $L_2$-error, plotted against (d) $H_{max}$ and (e) $z_0$.}
    \label{fig:bubble_large}
\end{figure}

Figure \ref{fig:bubble_large} presents the statistical performance of the various coordinate systems evaluated over a set of $N=500$ randomized topographies for the rising bubble test case. Panel (a) displays the histogram of the $L_2$-error distribution, revealing that while the Hybrid and SLEVE schemes exhibit similar performance profiles, the NEUVE coordinate successfully reduces the frequency of large error outliers. Analyzing performance sensitivity to terrain characteristics, we observe that the NEUVE coordinate maintains a lower absolute error across all maximum mountain heights tested (panel (b)). However, the relative improvement factor decreases as the mountain height increases (panel (d)). A comparable trend is observed with respect to topographic roughness (panels (c) and (e)). Although the overall performance gain of the NEUVE coordinate is more modest in this buoyancy-driven regime compared to the advection test, a consistent advantage remains evident.

\begin{figure}[h!]
    \centering
    \includegraphics[width=0.65\linewidth]{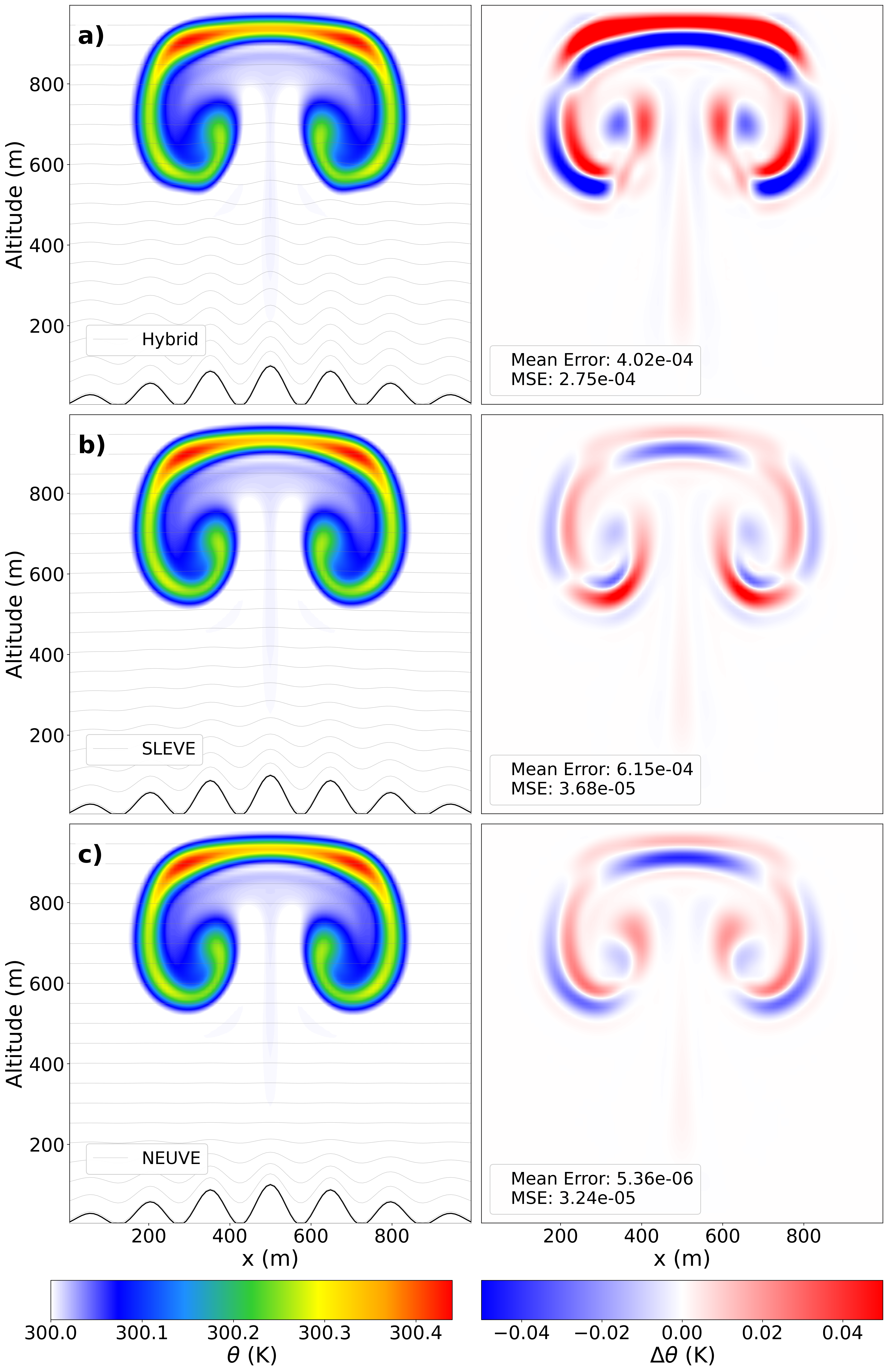}
    \caption{Rising thermal bubble test over a jagged mountain range. Same as Figure \ref{fig:bubble_large} but evaluated over high-frequency multi-peak topography ($h_c=100$ m, $a_c=400$ m, $\lambda_c=150$ m). The SLEVE and Hybrid parameters are tuned with $s_1 = 400$, $s_2 = 130$, and $s=400$.}
    \label{fig:bubble_complex}
\end{figure}

Figure \ref{fig:bubble_complex} shows an example of the solution and error where we sampled a more complex topography that remained stable for all formulations. In this regime, the analytic formulations struggle more than the simple bell-shaped mountain. The Hybrid coordinate exhibits an MSE of $2.75 \times 10^{-4}$, and the SLEVE coordinate shows a similar error with an MSE of $6.15 \times 10^{-4}$. In contrast, the NEUVE coordinate maintains high fidelity, with an MSE of $3.24 \times 10^{-5}$. This represents an improvement factor of approximately $\sim 19\times$ over SLEVE and $\sim 8.5\times$ over the Hybrid coordinate.

\subsection{Case 3: Density current}

The final test case involves a cold air bubble dropping in a neutral environment to form a density current \cite{Straka1993}. Unlike the rising bubble, the density current propagates along the lower boundary, interacting directly with the most strongly distorted coordinate surfaces near the terrain.

The domain is initialized with a constant potential temperature $\theta_0 = 300$ K. A cold perturbation is placed at $(x_c, z_c) = (0, 3000)$ m with an amplitude of $\Delta \theta_c = -15$ K:
\begin{equation}
    \Delta \theta(x,z) = 
    \begin{cases} 
      \frac{\Delta \theta_c}{2} \left[ 1 + \cos\left( \pi \tilde{r} \right) \right] & \text{for } \tilde{r} \le 1 \\
      0 & \text{otherwise},
    \end{cases}
\end{equation}
where $\tilde{r} = \sqrt{((x-x_c)/x_r)^2 + ((z-z_c)/z_r)^2}$ represents the normalized elliptic distance, with radii $x_r = 4000$ m and $z_r = 2000$ m. As the cold air descends and spreads laterally, the outflow boundary interacts with surface topography. This test is highly stringent regarding numerical stability; spurious vertical velocities generated by steep terrain slopes often lead to numerical instability at the sharp leading edge of the density current.

Since an analytic solution does not exist for this non-linear flow regime over complex topography, we rely on a qualitative analysis of numerical artifacts. Standard terrain-following coordinates frequently imprint the grid structure onto the flow solution, manifesting as spurious stationary vertical velocity waves above the mountain slopes. We therefore evaluate the performance of the coordinate systems based on their ability to suppress these unphysical artifacts and preserve the structural integrity of the density current front.

\begin{figure}[h!]
    \centering
    \includegraphics[width=0.8\linewidth]{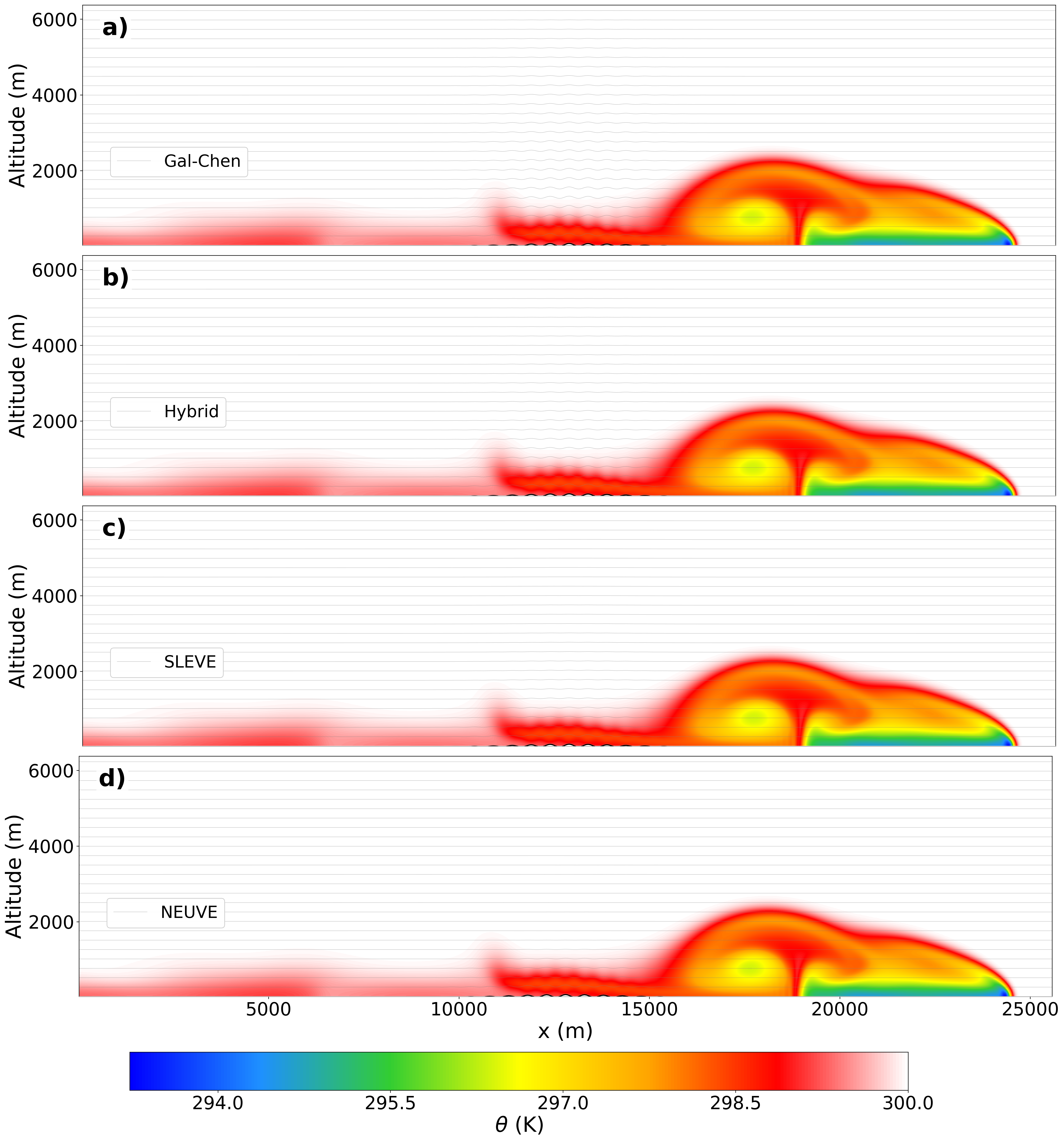}
    \caption{Density current interaction with small topography. Snapshot of potential temperature $\theta$ as the cold pool front traverses a mountain range defined by $h_c=75$ m, $a_c=2000$ m, $\lambda_c=500$ m. While the topography is small to ensure stability, coordinate artifacts are visible in the standard formulations (a-c) as numerical noise behind the front. The NEUVE coordinate (d) suppresses these artifacts, resulting in a cleaner representation of the cold pool body. The SLEVE and Hybrid parameters are tuned with $s_1 = 300$, $s_2 = 1000$, and $s=3000$}
    \label{fig:density_current}
\end{figure}

\begin{figure}[h!]
    \centering
    \includegraphics[width=0.8\linewidth]{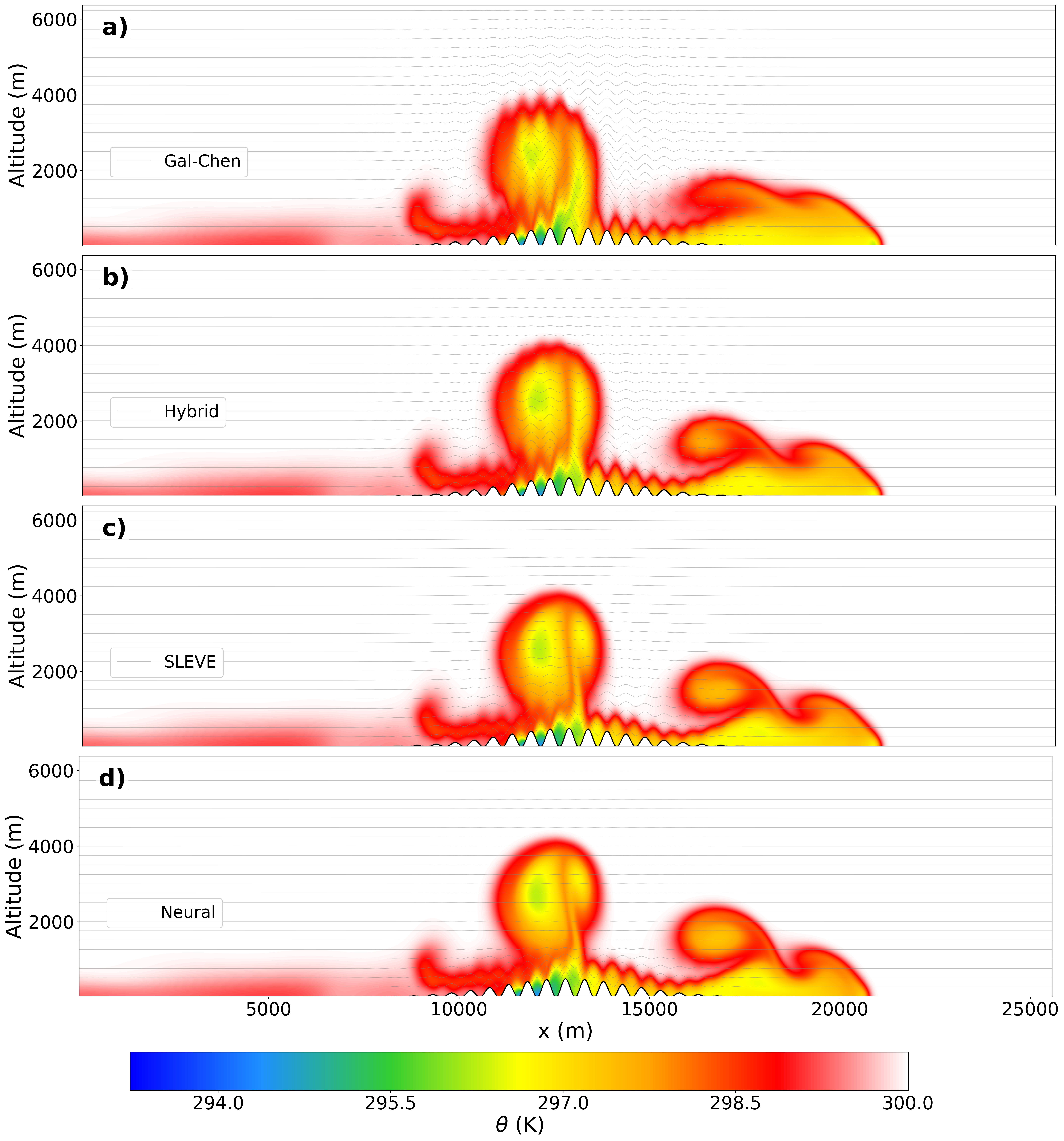}
    \caption{Density current interaction with steep jagged topography. Snapshot of potential temperature $\theta$ over a larger, steeper mountain range ($h_c=500$ m, $a_c=2500$ m, $\lambda_c=500$ m). (a-c) The Gal-Chen and Somerville, Hybrid, and SLEVE coordinates generate spurious vertical velocity modes (visible as vertical striations above the peaks). (d) The NEUVE coordinate significantly reduces these stationary grid imprints. The SLEVE and Hybrid parameters are tuned with $s_1 = 300$, $s_2 = 1000$, and $s=3000$}
    \label{fig:density_current_hyper}
\end{figure}

Figures \ref{fig:density_current} and \ref{fig:density_current_hyper} compare the vertical velocity fields ($w$) as the current passes over the mountain peaks. In the standard Gal-Chen and Somerville, Hybrid, and SLEVE coordinate solutions, we observe distinct stationary artifacts in the vertical velocity field. These manifest as vertical striations or standing waves extending upward from the mountain slopes. These artifacts are physically spurious; they arise because the horizontal flow is erroneously forced upward by the oscillating coordinate surfaces. 
The NEUVE coordinate significantly suppresses these stationary grid imprints. As seen in the bottom panels of Figure \ref{fig:density_current_hyper}, the vertical velocity field is notably smoother, with the spurious striations largely eliminated. 

Beyond the vertical velocity artifacts, the choice of coordinate system influences the structural evolution of the density current as it propagates past the mountain range. In the Gal-Chen and Somerville, and Hybrid simulations (Figures \ref{fig:density_current_hyper}a, b), the interaction between the flow and the distorted grid generates numerical noise that physically diffuses the interface between the cold pool and the environment. This manifests as a degradation of the density current head and a blurring of the Kelvin--Helmholtz billows forming in the wake. In contrast, the NEUVE coordinate solution (Figure \ref{fig:density_current_hyper}d) preserves the coherent structure of the propagating front.

\section{Discussion}
\label{Sec:Discussion}

The results presented demonstrate that differentiable programming offers a viable pathway to overcome the rigid, heuristic design of traditional vertical coordinates. A central finding is the generalization capability of the NEUVE coordinate transformation across disparate physical regimes. The NEUVE coordinate was trained exclusively on a linear passive advection problem (Case 1) with stochastic topography and was never exposed to the non-linear Euler equations, buoyancy forces, or the steep gradients associated with density currents during the optimization phase. Despite this limited training, the NEUVE coordinate successfully generalized to the complex dynamics of the rising bubble (Case 2) and density current (Case 3). This suggests that the network did not merely overfit to the proxy task, but rather learned geometric properties of the grid that are universally beneficial for the finite-difference discretization.

While the NEUVE coordinate demonstrates robustness across these non-linear test cases, the grid structure exhibits a clear adaptation to the specific vertical profile of the training data, most notably the transition to a flat Cartesian grid at the base of the constant flow layer ($z \approx 5$~km). This sensitivity indicates a capability for regime-specific optimization. Unlike analytic coordinates that rely on heuristic scale heights fixed a priori for all atmospheric states, our framework allows the discretization to be objectively ``tuned'' to the dominant vertical structure of a specific region or flow regime. For operational contexts with persistent features, such as trade wind inversions or low-level jets, the coordinate system can be retrained to minimize discretization errors specifically in those critical layers.

The practical implications of the learned grid can be quantified with respect to the numerical convergence analysis demonstrated in the Appendix \ref{APP:Advection_Convergence}. Figure \ref{fig:convergence} demonstrates how the NEUVE coordinate maintains a consistent error reduction across all resolutions tested. Notably, to achieve an error magnitude equivalent to the NEUVE coordinate at a coarse resolution of $\Delta x= 1000$~m, the standard Gal-Chen and Somerville coordinate requires a resolution refinement to approximately $\Delta x \approx 250$~m. This shows that the optimized coordinate allows for a four-fold coarsening of the horizontal grid without compromising accuracy. In a three-dimensional operational context, such a reduction in required grid points could translate to orders-of-magnitude savings in computational cost.

A distinct methodological contribution of this work, which complements the learned grid geometry, is the calculation of geometric metric terms via automatic differentiation. In standard terrain-following models, coordinate derivatives (e.g., $\partial z / \partial \xi$) are typically approximated using finite differences, introducing geometric truncation errors that can alias into the pressure gradient force. By leveraging the differentiable framework, we compute these terms exactly to machine precision. This eliminates a source of inconsistency over steep slopes, offering a numerical advantage that exists independently of the neural parameterization.

The framework presented here is naturally extensible to three-dimensional atmospheric models. Since terrain-following coordinates typically operate on independent vertical columns, the learned 2D decay profile can be applied column-wise to 3D topography without modification. The integration of neural networks into the solver imposes a negligible overhead for such static topography configurations. Additionally, the implementation in \texttt{JAX} leverages XLA to compile the entire dynamical core including the exact metric computations into optimized kernels for GPU or TPU execution.

A limitation of the current framework is its reliance on supervised training against high-fidelity analytic reference solutions. This constraint presently restricts the optimization phase to idealized test cases where exact solutions are known, such as linear advection or hydrostatic mountain waves. To extend this methodology to fully turbulent or varying flow regimes where no analytic solution exists, future work could explore unsupervised loss functions. A promising avenue is to directly minimize the theoretical discretization error estimates derived by \citeA{ANewTerrainFollowingVerticalCoordinateFormulationforAtmosphericPredictionModels} or~\citeA{huang2010adaptive} for curvilinear meshes.

Purely data-driven models benefit from native efficiency on hardware accelerators and automated fine-tuning via backpropagation. In contrast, operational numerical weather prediction systems often rely on legacy codebases where algorithmic choices are manually tuned and structurally rigid. Current hybrid modeling efforts have largely addressed this by learning subgrid-scale parameterizations, yet they leave the numerical backbone of the solver untouched. Our results suggest that ML can be effectively applied to this numerical discretization itself. We demonstrate that structural components such as the vertical coordinate are just as amenable to gradient-based optimization as physical closures.

This opens a pathway for the incremental modernization of dynamical cores. By successively replacing classical heuristic components with learned equivalents, it is possible to transition towards tailored combined physics and data-driven systems in a constructive, iterative manner. Rather than constructing a fully data-driven model (such as GraphCast or Pangu Weather) from scratch one can choose which model component to replace using ML, as well as how machine learned components can be integrated back into a classical dynamical core. This is exemplified by our machine learned vertical coordinate NEUVE could be transferred into existing classical numerical weather prediction models. In the limit, this process could also lead to a fully learned model achieved through the gradual replacement of individual components rather than a complete discard of the physical solver (an ``ML Ship of Theseus''). By establishing this framework for a fully differentiable RCM, we open the door to automating other structural components, such as lateral boundary conditions or transport dynamics \cite{taylor2025diffeomorphicneuraloperatorlearning}. For instance, a differentiable model could learn relaxation strategies that minimize the variable ``spatial spin-up'' distance identified by \citeA{gmd-17-1497-2024}, thereby reducing computational overhead. Beyond parameter tuning, this architecture enables new forms of physical analysis: \citeA{Bano-Medina2025} leverage exact gradients to compute flow sensitivities for cyclone dynamics, while \citeA{whittaker2025constructingextremeheatwavestorylines} utilize the framework to efficiently sample rare extremes via minimal state perturbations.

\section{Conclusions}
\label{Sec:Conclusions}
In this work, we presented a framework to optimize the vertical coordinate discretization of atmospheric models using differentiable programming. By treating the vertical coordinate transformation as a learnable component within a differentiable 2D non-hydrostatic dynamical core, we successfully replaced heuristic analytic decay functions with a neural network parameterization optimized via gradient descent.

Our results demonstrate that a coordinate system trained on a simplified linear advection task effectively generalizes to complex, non-linear flow regimes. In benchmark tests involving rising thermal bubbles and density currents over steep topography, the NEUVE coordinate significantly reduced discretization errors and suppressed the spurious vertical velocity artifacts common in standard terrain-following formulations.

%
%
\section*{Acknowledgments}

This work was supported in part by the Canada Research Chairs program and the Natural Sciences and Engineering Research Council of Canada (NSERC) Discovery Grant program. T.W.\ acknowledges support from an NSERC Canada Graduate Scholarship – Doctoral (CGS D) (Grant No. 588387-2024). S.T.\ acknowledges financial support from the Pacific Institute for the Mathematical Sciences and Environment and Climate Change Canada. The authors would like to thank Eleanor McGinn for helpful comments and suggestions on the clarity and presentation of the manuscript.

\section*{Data availability statement}
All the data can be generated using the code which will be made publicly available upon publication.


%
%

\bibliographystyle{apacite}
\bibliography{agusample}

\appendix 

\section{Governing equations in generalized vertical coordinates}
\label{app:derivation}

In this section we provide a derivation of the transformed Euler equations using a generalized vertical coordinate. The derivation follows~\citeA{coiffier2011fundamentals}, and further information on the formulation of numerical schemes in adaptive coordinates can be found in~\citeA{huang2010adaptive}. The governing equations are transformed from the physical to the computational domain through the coordinate transformation $\varphi(\xi,\zeta) = (x, z(\xi,\zeta))$. The horizontal coordinate $x=\xi$ remains fixed, while the terrain-following coordinate $z=z(\xi, \zeta)$ is defined by the topography and the vertical decay function $B(\zeta)$ given in Section~\ref{subsec:VerticalCoodinates}. Using the inverse Jacobian of this transformation, one can derive the transformation rules for the spatial derivative operators defined by
\begin{equation}\label{eq:TransformedPartialDerivatives}
\frac{\partial}{\partial z} = \frac{1}{z_{\zeta}}\frac{\partial}{\partial \zeta} \,,  \quad \frac{\partial}{\partial x} = \frac{\partial}{\partial \xi} - \frac{z_\xi}{z_{\zeta}} \frac{\partial}{\partial \zeta},
\end{equation}
for the vertical derivative and horizontal derivatives where we use the shorthand $z_{\zeta}$ and $z_{\xi}$ to denote the coordinates' partial derivatives. We note that these formulas highlight the necessity of monotonicity of the coordinate $z$, which guarantees that $\det(D\varphi) = J>0$. The expressions of the divergence of a vector field $\mathbf{u}$, the directional derivative of a scalar field $\mu$, and the Laplacian of a scalar function become
\begin{equation}
\label{transformed_operators}
\begin{aligned}
    \nabla \cdot \mathbf{u} &= \frac{\partial u}{\partial \xi} + \frac{1}{z_\zeta} \left( \frac{\partial w}{\partial \zeta} - z_\xi \frac{\partial u}{\partial \zeta} \right),
    \\
    \mathbf{u} \cdot \nabla f &= u \left( \frac{\partial f}{\partial \xi} - \frac{z_\xi}{z_\zeta}\frac{\partial f}{\partial \zeta} \right) + w \left( \frac{1}{z_\zeta} \frac{\partial f}{\partial \zeta} \right)\,,
    \\
    \Delta f & = \frac{\partial^2 f}{\partial \xi^2} -2\frac{z_{\xi}}{z_{\zeta}} \frac{\partial^2 f}{\partial\xi \partial \zeta} + \frac{1 + z_{\xi}^2}{z_{\zeta}^2}\frac{\partial^2 f}{\partial \zeta^2} - \frac{1}{z_{\zeta}^3}\big(z_{\xi\xi}z_{\zeta}^2 - 2z_{\xi}z_{\xi\zeta}z_{\zeta} + (1 + z_{\xi})^2 z_{\zeta\zeta}\big)\frac{\partial f}{\partial \zeta} \,.
\end{aligned}
\end{equation} 
Expressing the state variable $(\boldsymbol{u}, \pi, \theta)(x,z) = (\boldsymbol{u}, \pi, \theta)\circ \varphi(\xi, \zeta)$, the non-conservative form of the Euler equations in generalized coordinates vertical coordinates can be written as
\begin{equation}
\label{eq:EulerEquationsTransformed}
\begin{aligned}    
    \frac{\partial \pi}{\partial t} + \left[ u \frac{\partial \pi}{\partial \xi} + \frac{1}{z_\zeta}(w - u z_\xi)\frac{\partial \pi}{\partial \zeta} \right] + \frac{R}{c_v} \pi \left(\frac{\partial u}{\partial \xi} + \frac{1}{z_\zeta} \left( \frac{\partial w}{\partial \zeta} - z_\xi \frac{\partial u}{\partial \zeta} \right)\right) &= 0
    \\
    \frac{\partial u}{\partial t} + u \left( \frac{\partial u}{\partial \xi} - \frac{z_\xi}{z_\zeta}\frac{\partial u}{\partial \zeta} \right) + w \left( \frac{1}{z_\zeta} \frac{\partial u}{\partial \zeta} \right) + c_p \theta \left( \frac{\partial \pi}{\partial \xi} - \frac{z_\xi}{z_\zeta}\frac{\partial \pi}{\partial \zeta} \right) &= \nu \Delta u
    \\
    \frac{\partial w}{\partial t} + u \left( \frac{\partial w}{\partial \xi} - \frac{z_\xi}{z_\zeta}\frac{\partial w}{\partial \zeta} \right) + w \left( \frac{1}{z_\zeta} \frac{\partial w}{\partial \zeta} \right)+ c_p \theta \left( \frac{1}{z_\zeta}\frac{\partial \pi}{\partial \zeta} \right) &= -g + \nu \Delta w,
    \\
    \frac{\partial \theta}{\partial t} + u \left( \frac{\partial \theta}{\partial \xi} - \frac{z_\xi}{z_\zeta}\frac{\partial \theta}{\partial \zeta} \right) + w \left( \frac{1}{z_\zeta} \frac{\partial \theta}{\partial \zeta} \right) &= 0.
\end{aligned}
\end{equation}
where $\Delta u, \Delta w$ are computed following \eqref{transformed_operators}. Physical diffusion is required for the density current test case, cf.~\citeA{GIRALDO20083849}. 

\section{Numerical discretization}
\label{app:numerics}

The system of equations~\eqref{eq:EulerEquationsTransformed} is discretized on a staggered Arakawa C-grid in the computational coordinates. The prognostic state vector is $\mathbf{q} = (u, w, \pi, \theta)^{\rm T}$, where the scalar variables $\pi$ and $\theta$ are defined at cell centers $(i,k)$, the horizontal velocity $u$ is defined at vertical cell faces $(i+\tfrac{1}{2},k)$ and the vertical velocity $w$ is defined at horizontal cell faces $(i, k+\tfrac{1}{2})$. A key feature of our solver is the exact calculation of metric terms $z_{\xi}$ and $z_{\zeta}$ using \texttt{JAX}'s automatic differentiation engine. Unlike standard formulations that discretize the analytic coordinate transformation using finite differences, our approach computes these derivatives to machine precision. This ensures that the geometric conservation law is satisfied and eliminates the truncation errors associated with metric computation that can otherwise lead to spurious forcing over steep topography~\cite{klemp2003numerical}. We employ second-order centerd finite differences for all advective fluxes which are multiplied by the automatically differentiated coordinate transformations. For the test cases requiring physical viscosity, thus effectively solving the Navier--Stokes instead of the Euler equations, we implement the full Laplacian operator in transformed physical coordinates in flux form. In our implementation, we calculate the gradients on the staggered cell faces, with cross-derivative terms handled via averaging, and compute the divergence of these fluxes again at the cell centers.\par

The boundary conditions are enforced in two stages; implicitly through the operator stencils and explicitly through state variable constraints. At the model top, $z=H$, we enforce a rigid lid via setting $w=0$. At the surface, $z=z_s$, we enforce the kinematic boundary condition: for flat terrain, $w=0$ while for cases with variable topography, we enforce tangent flow $w = u \partial z_s / \partial x$.
For the advection test case periodic boundaries are used, although they have no impact as the flow simulation is short enough that the cosine bell does not transition through the boundary. For the bubble and density current benchmarks we impose free-slip solid wall lateral boundary conditions. The horizontal velocity is set to $u$ at the domain edges and scalar gradients are set to zero.

To prevent wave reflection at the boundaries in non-periodic domains, we apply a Rayleigh damping sponge layer to the velocity and potential temperature fields. The damping coefficient $\tau(x,z)$ ramps up using a squared-sine profile over the outer 15\% of the domain width and top 25\% of the domain height. To stabilize the explicit time stepping (see below), and to filter out grid-scale noise, we apply standard fourth-order hyperdiffusion. Following standard practice, this operator is computed in computational coordinates rather than in physical coordinates, to avoid the computation of the expensive cross-terms arising due to the form of the transformed spatial derivatives~\eqref{eq:TransformedPartialDerivatives}, while still being effective at suppressing small $\mathcal O(\Delta x)$ noise modes with $\nu_4$ chosen to guarantee stability of the numerical simulations over the time horizon of integration. \par

Time integration is performed using a third-order Strong Stability Preserving Runge-Kutta (SSPRK3) scheme given by
\begin{align*}
    \mathbf{q}^{(1)} &= \mathbf{q}^n + \Delta t L(\mathbf{q}^n) \\
    \mathbf{q}^{(2)} &= \frac{3}{4}\mathbf{q}^n + \frac{1}{4}\mathbf{q}^{(1)} + \frac{1}{4}\Delta t L(\mathbf{q}^{(1)}) \\
    \mathbf{q}^{n+1} &= \frac{1}{3}\mathbf{q}^n + \frac{2}{3}\mathbf{q}^{(2)} + \frac{2}{3}\Delta t L(\mathbf{q}^{(2)}).
\end{align*}
Explicit boundary conditions are re-applied after the computation of each stage.

\subsection{Numerical validation}
\label{APP:Advection_Convergence}

\begin{figure}[h!]
    \centering
    \includegraphics[width=0.5\linewidth]{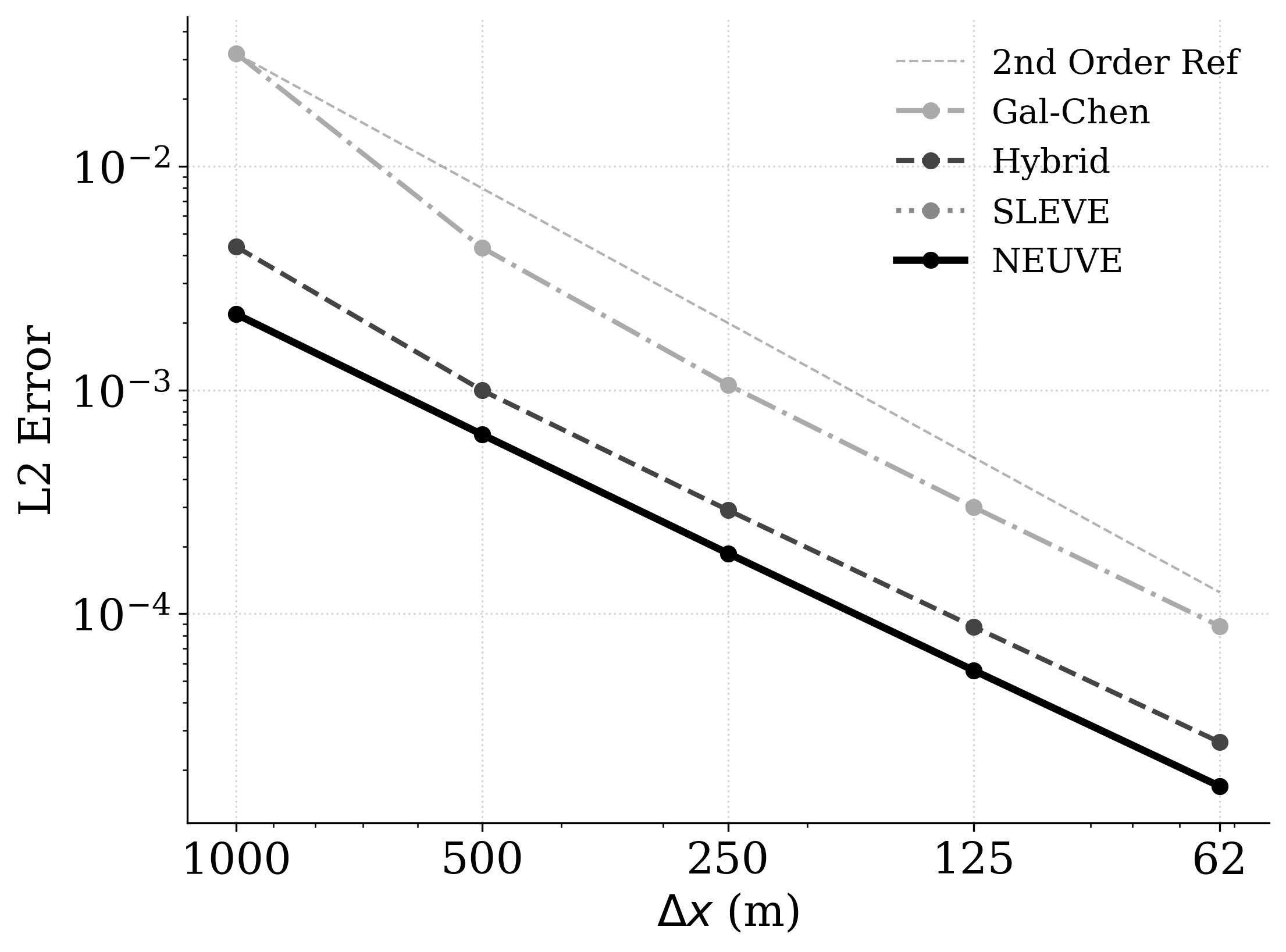}
    \caption{Grid convergence study for the passive scalar advection test.}
    \label{fig:convergence}
\end{figure}

To validate the numerical implementation and quantify the accuracy gains of the learned coordinate system, we performed a grid convergence study using the linear advection benchmark (Case 1). The simulations were integrated for $T=5000$ s over a static topography profile. We varied the horizontal resolution from $\Delta x = 1000$ m to $\Delta x = 62.5$ m, scaling the vertical resolution and time step proportionally to maintain a constant Courant--Friedrichs--Lewy (CFL) number. Figure~\ref{fig:convergence} displays the $L_2$-error norm of the passive scalar field at the final time $T$. 

All coordinate formulations (Gal-Chen and Somerville, Hybrid, SLEVE, and NEUVE) exhibit error convergence rates parallel to the theoretical second-order reference line. This confirms that the numerical discretization maintains its designed second-order accuracy regardless of the vertical coordinate transformation employed.

While the convergence rates are identical, the NEUVE coordinate shows a consistent reduction in the absolute error magnitude. As noted in the discussion, the NEUVE coordinate at a coarse resolution of $\Delta x = 1000$ m achieves an error magnitude comparable to the standard Gal-Chen coordinate at $\Delta x \approx 250$ m. This demonstrates that the optimized grid structure allows for a roughly four-fold coarsening of the grid and a commensurate reduction in computational cost without compromising solution fidelity for this flow regime.

\section{Hyperparameter sweep}
\label{APP:Hyper}

\begin{figure}[h!]
    \centering    \includegraphics[width=0.5\linewidth]{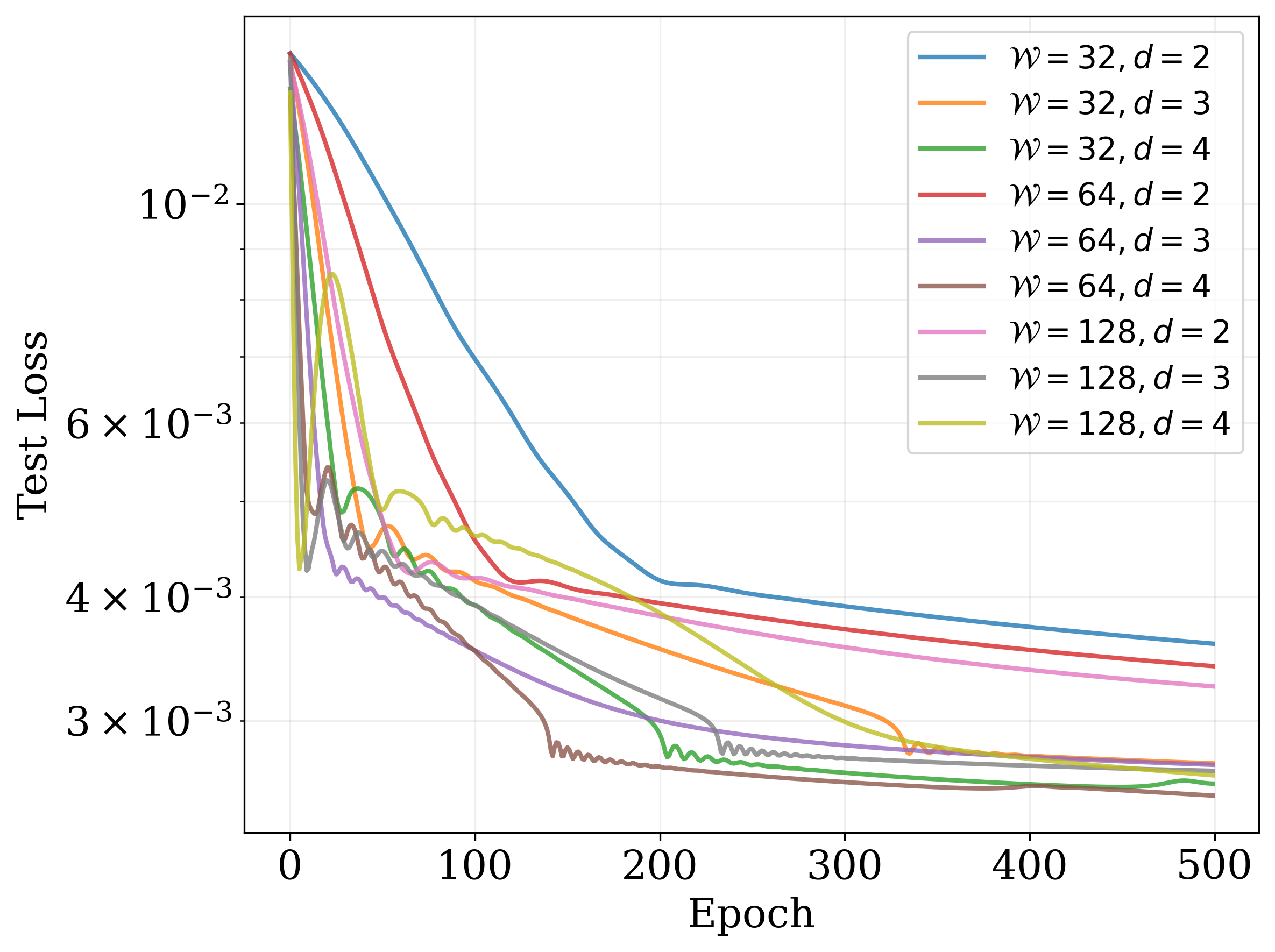}
    \caption{Training dynamics for the NEUVE coordinate hyperparameter sweep showing the validation loss history over 500 epochs for varying network widths ($\mathcal{W}$) and depths ($d$).}    \label{fig:sweep_history}
\end{figure}

To determine the optimal architecture for the NEUVE coordinate parameterization, we performed a hyperparameter grid search over the network depth ($d \in \{2, 3, 4\}$) and width ($\mathcal{W} \in \{32, 64, 128\}$). All candidates were trained on the linear advection task (Case 1) using the stochastic topography protocol described in Section \ref{subsec:optimization}. Figure \ref{fig:sweep_history} presents the loss history on a test set for the sweep. The results reveal that increasing the layer width provides the most consistent improvement in performance; the narrowest networks ($\mathcal{W}=32$) converge slowly and plateau at a higher loss compared to the wider configurations. While increasing the width to $\mathcal{W}=128$ yields the lowest absolute error, the marginal gain over $\mathcal{W}=64$ diminishes relative to the increased computational cost of the forward pass. Regarding network depth, the deeper models ($d=4$) exhibit greater volatility during the initial training phase (epochs 0--50), characterized by large oscillations in the loss function, whereas the shallower networks are more stable but plateau earlier. Based on this analysis, we selected the configuration with 3 layers of 64 units ($\mathcal{W}=64, d=3$) for the coordinate. We do not perform a parameter sweep on the other parameters, they are fixed and presented in Table~\ref{train_details}.

\begin{table*}[!ht]
\center
\begin{tabular}{@{}lrrr@{}}
\hline
learning rate & batch size & Number of epochs & Gradient Clipping\\ \hline
$10^{-3}$ & 30 & 500 & 1.0 \\
\hline
\end{tabular}
\caption{Parameters used to train the NEUVE coordinates.}
\label{train_details}
\end{table*}

\section{Experiment parameters}
\label{APP:Exp_params}
Table~\ref{sim_details} contains the parameters chosen for each of the numerical simulations carried out in this work.

\begin{table*}[!ht]
\resizebox{1.\linewidth}{!}{
\ra{1.0}
\begin{tabular}{@{}lrrrrrrrr@{}}
\hline
Simulation name
& $L_x$ & $L_z$ & $T$
& $\Delta x$ & $\Delta z$
& $\Delta t$
& $\nu$ & $\nu_4$ \\ \hline

Advection (Fig.~\ref{fig:advect_stats} \& \ref{fig:advection_test})
& 300000 & 25000 & 5000
& 500 & 250
& 12
& 0 & 0 \\

Bubble Statistical (Fig.~\ref{fig:bubble_large})
& 1000 & 1000 & 700
& 4.95 & 4.95
& 0.0025
& 0 & 200 \\

Bubble 2 (Fig.~\ref{fig:bubble_complex})
& 1000 & 1000 & 700
& 9.90 & 9.90
& 0.005
& 0 & 200 \\

Density current 1 (Fig.~\ref{fig:density_current})
& 25600 & 6400 & 1800
& 49.9 & 49.6
& 0.05
& 75 & 0 \\

Density current 2 (Fig.~\ref{fig:density_current_hyper})
& 25600 & 6400 & 1800
& 24.9 & 24.8
& 0.025
& 75 & 0 \\

\hline
\end{tabular}
}
\caption{Simulation configurations used in this study. All units are SI. Grid spacings are defined as $\Delta x = L_x/N_x$ and $\Delta z = L_z/N_z$.}
\label{sim_details}
\end{table*}

\section{Application to analytical coordinate tuning}
\label{app:param_opt}
\begin{figure}[h!]
    \centering
    \includegraphics[width=0.75\linewidth]{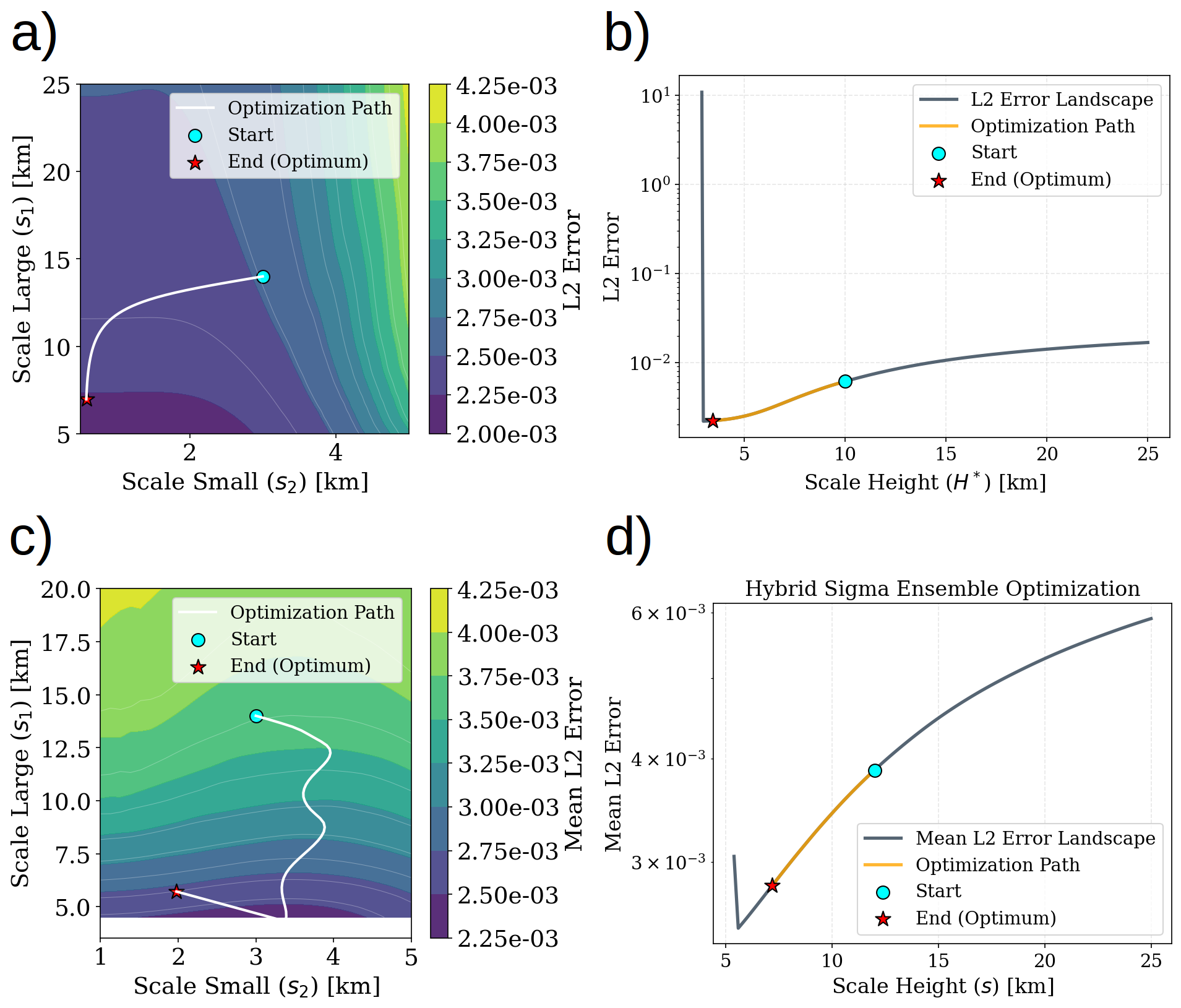}
    \caption{Loss landscape and optimization trajectory for the SLEVE and Hybrid parameters on a single topography realization. (a) Contour map of the simulation $L_2$-error as a function of the large-scale ($s_1$) and small-scale ($s_2$) decay parameters. The white line traces the path of the gradient-based optimization from initialization (cyan circle) to the converged optimum (red star). (b) Profile of the error landscape with respect to the scale height $s$, showing the wall of instability (sharp vertical asymptote) at low scale heights where grid orthogonality is violated. Panels c) and d) are the SLEVE and Hybrid parameters trained an on ensemble of 30 topographic members.}
    \label{fig:opt_coord}
\end{figure}

In addition to tuning the NEUVE coordinates, the differentiable programming scheme allows us to also consider tuning the parameters of existing analytic formulations. Figure \ref{fig:opt_coord}a visualizes the optimization of the scalar scale-height parameters ($s_1, s_2$) for the SLEVE coordinate on a single topography instance. Figure \ref{fig:opt_coord}a shows the loss landscape for the $s$ parameter of the hybrid coordinates. The loss landscape reveals a distinct ``wall of instability'', representing parameter combinations that induce grid tangling or numerical divergence. The gradient-based optimization trajectory (white line) successfully navigates this constraint, converging from a suboptimal initialization to a minimum that balances grid smoothness with terrain-following requirements. While we show that it is possible to tune analytic coordinates to specific flow regimes, the remainder of this work utilizes the standard parameter values from \citeA{ANewTerrainFollowingVerticalCoordinateFormulationforAtmosphericPredictionModels}, scaled to the experimental domain. Figures \ref{fig:opt_coord}c and \ref{fig:opt_coord}d, show the loss landscape when considering training on an ensemble of topographies. In this case we can see how when we let the training reach an unstable region, the regulation term pushes the optimized values back into a stable region. The ensemble training aims for reducing the mean error over any topography and seems to lead to similar values as the single trajectory training scheme.

\end{document}